%% file: main.tex
\documentclass[10pt]{article}

\usepackage{fullpage}
\usepackage{setspace}
\usepackage{parskip}
\usepackage{titlesec}
\usepackage[section]{placeins}
\usepackage{xcolor}
\usepackage{breakcites}
\usepackage{lineno}
\usepackage{hyphenat}

\PassOptionsToPackage{hyphens}{url}
\usepackage[colorlinks = true,
            linkcolor = blue,
            urlcolor  = blue,
            citecolor = blue,
            anchorcolor = blue]{hyperref}
\usepackage{etoolbox}

\renewenvironment{abstract}
  {{\bfseries\noindent{\abstractname}\par\nobreak}\footnotesize}
  {\bigskip}

\titlespacing{\section}{0pt}{*3}{*1}
\titlespacing{\subsection}{0pt}{*2}{*0.5}
\titlespacing{\subsubsection}{0pt}{*1.5}{0pt}

\usepackage{authblk}

\usepackage{graphicx}
\usepackage[space]{grffile}
\usepackage{latexsym}
\usepackage{textcomp}
\usepackage{longtable}
\usepackage{tabulary}
\usepackage{booktabs,array,multirow}
\usepackage{amsfonts,amsmath,amssymb}
\providecommand\citet{\cite}
\providecommand\citep{\cite}

\newif\iflatexml\latexmlfalse

\AtBeginDocument{\DeclareGraphicsExtensions{.pdf,.PDF,.eps,.EPS,.png,.PNG,.tif,.TIF,.jpg,.JPG,.jpeg,.JPEG}}

\usepackage[utf8]{inputenc}
\usepackage[english]{babel}

\usepackage{float}

\begin{document}

\title{Waveform Simulation in PandaX-4T}


\input{authorlist}


\vspace{-1em}

  \date{\today}

\begingroup
\let\center\flushleft
\let\endcenter\endflushleft
\maketitle
\endgroup


\begin{abstract}
Signal reconstruction through software processing is a crucial component of the  background and signal models in the PandaX-4T experiment, which is a multi-tonne dark matter direct search experiment. 
The accuracy of signal reconstruction is influenced by various detector artifacts, including noise, dark count of photomultiplier, impurity photoionization in the detector, and other relevant considerations.
In this study, we present a detailed description of a semi-data-driven approach designed to simulate the signal waveform. 
This work provides a reliable model for the efficiency and bias of the signal reconstruction in the data analysis of PandaX-4T. 
By comparing critical variables which relate to the temporal shape and hit pattern of the signals, we demonstrate a good agreement between the simulation and data.
\end{abstract}

\maketitle

\sloppy

\section{Introduction}

The search for dark matter (DM)~\cite{bertone2005particle, jungman1996supersymmetric, feng2010dark} is a highly active frontier in astroparticle physics. 
Among the various experimental approaches~\cite{xiao2015low, tan2016dark, collaboration2018dark, akerib2017results, agnese2018results, petricca2020first, meng2021dark, aprile2023first, lux2023first}, dual-phase xenon time projection chamber (TPC) based experiments have been recognized as highly sensitive for detecting cold DM particles within the mass range of approximately 10\,GeV/c$^2$ to TeV/c$^2$.
PandaX-4T~\cite{meng2021dark}, located in the China Jinping Underground Laboratory (CJPL)~\cite{kang2010status, cheng2017china}, is a typical experiment of this kind.
PandaX-4T employs a cylindrical dual-phase xenon TPC with a sensitive volume measuring approximately 1.2 meters in diameter and 1.2 meters in height.
The TPC is equipped with 368 Photomultipliers (PMTs), 169 of which are distributed at the top PMT array and 199 at the bottom.
The TPC detects and measures both the prompt scintillation signal ($S1$) and the subsequent amplified proportional scintillation signal ($S2$), which arises from the delayed ionization signals.
By analyzing the time difference between $S1$ and $S2$ signals, as well as the spatial distribution of PMT hits associated with the $S2$ signal, we are capable of reconstructing the vertical and horizontal positions of the interaction vertex, respectively.
Accurate 3-D position reconstruction, along with the determination of the $S2/S1$ ratio, plays a crucial role in particle discrimination within the TPC detector. 
This discrimination capability is of crucial importance as it greatly reduces the effective background for DM direct searches.

In the interpretation of the DM search results in PandaX-4T, precise models for the low-energy background and the DM signal are of utmost importance. 
A critical component of the low-energy models is the reconstruction of the $S1$ and $S2$.
This reconstruction process involves various steps such as peak identification, pulse classification, clustering, and other relevant techniques.
In order to accurately evaluate the efficiency and potential biases inherent in the software reconstruction, it is essential to obtain pure  events that faithfully represent the desired physical signal. 
However, acquiring such pure samples from the recorded reconstructed data is challenging, as the recorded data have already been influenced by the effects of the software signal reconstruction inefficiency and biases.
To address this challenge, PandaX-4T has developed a dedicated waveform simulation (WS) framework that generates synthetic data waveforms. 
The WS framework incorporates our best understanding of the processes involved in the generation, collection, and reconstruction of $S1$ and $S2$ signals, as well as the accompanying sources of noise, such as dark counts, PMT afterpulsing, and impurity photoionization.
In addition, the extensive samples generated by the WS can also be utilized to train machine learning and deep neural network algorithms, which can further enhance background rejection capabilities.

This manuscript presents a detailed account of the WS employed in various scientific studies conducted using the PandaX-4T data, including the search for Weakly Interacting Massive Particles (WIMPs)~\cite{meng2021dark}, the detection of solar $^8$B neutrinos~\cite{ma2023search}, investigations into light dark matter particles~\cite{li2023search}, and other related research endeavors~\cite{gu2022first, zhang2022search}.
Section~\ref{sec:waveform_simulation} of the manuscript elucidates the details of simulating the $S1$ and $S2$ pulses, as well as incorporating various sources of noise into the simulation process. 
Notably, the comparison between simulation and real data is presented in Section~\ref{sec:compare_to_data}, wherein critical variables such as signal width, pattern, and charge are evaluated.
Following the comparison, the manuscript concludes by summarizing the work and offering a discussion on its future perspectives.
Similar works of other experiments can be found in Ref.~\cite{XENONnT_WFSim}.

\section{Waveform simulation}
\label{sec:waveform_simulation}

\begin{figure}[htp]
    \centering
    \includegraphics[width=0.8\textwidth]{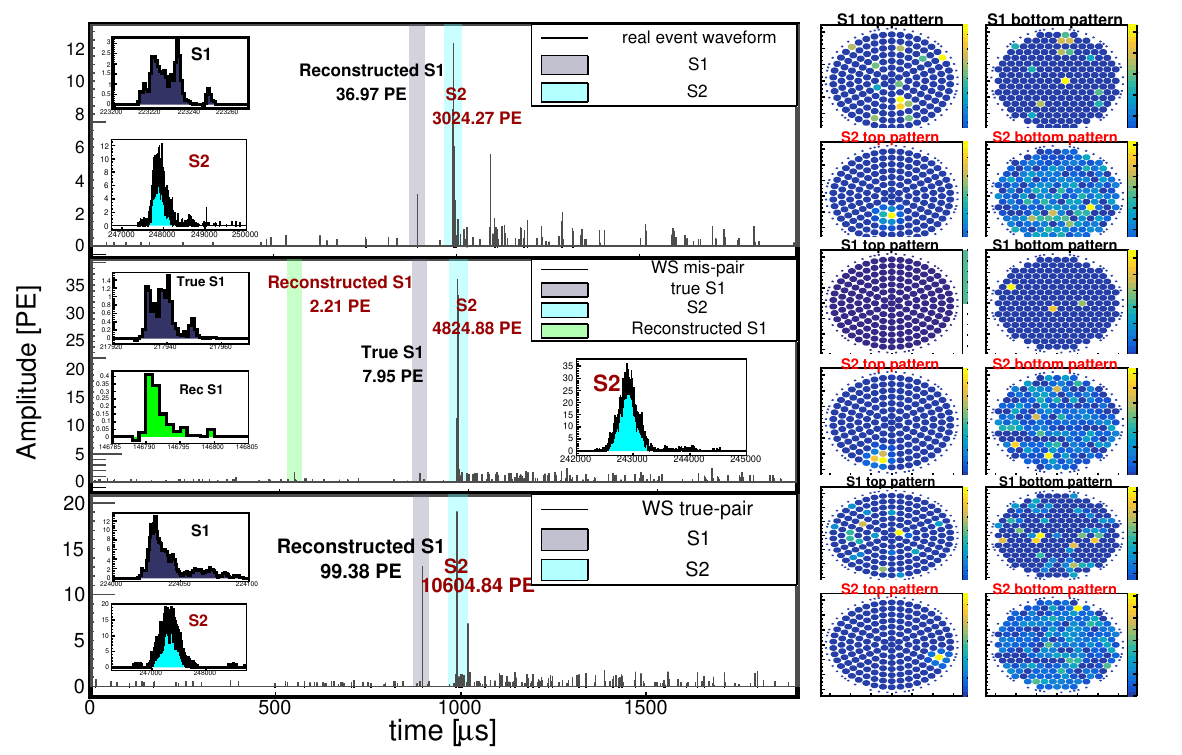}
    \caption{
    Three waveforms corresponding to events with similar drift times are depicted. 
    The top panel represents a recorded data waveform. 
    In the middle panel, a simulated waveform is shown where the true $S1$ signal is overshadowed by a noise $S1$ signal with a slightly higher charge. 
    The bottom panel displays a simulated waveform where both the $S1$ and $S2$ signals are correctly identified.
    The gray, green, and cyan shaded regions represent the time windows associated with the true $S1$, misidentified $S1$, and true $S2$ signals, respectively. 
    The reconstructed charge of the identified $S1$s and $S2$s are displayed in the panels.
    For enhanced detail, the inset panels provide zoomed-in views of the $S1$ and $S2$ signals.
    The hit patterns of the main $S1$s and $S2$s are also shown in the right columns.
    }
    \label{fig:example_sim_wfs}
\end{figure}

The simulated event waveforms in WS further undergo the software processing and reconstruction chain that is used for data processing and analysis in PandaX-4T, to ensure that the efficiency and bias of the simulated waveforms align with the observed data.
The WS employs a semi-data-driven approach to simulate the $S1$ and $S2$ waveforms. 
It takes into account various spurious pulses, including noise, dark counts, PMT afterpulsing, and delayed afterglow after a large $S2$ arising from impurity ionization and electron train effects. 
To illustrate, several simulated waveforms are presented in Fig.~\ref{fig:example_sim_wfs}.
Among these examples, one demonstrates successful identification of both the $S1$ and $S2$ signals, albeit with a slight difference in the reconstructed charge values with respect to the true charge, due to fluctuation in the reconstruction steps and interference from spurious pulses. 
In contrast, another example exhibits complete misidentification of the $S1$ signal, indicating a potential inefficiency in our software signal reconstruction process.

\subsection{S1 pulse}


The $S1$ signal represents the prompt scintillation signals detected in the TPC detector, arising from the interaction of particles with the xenon shell electrons or nucleus. 
The $S1$ signal exhibits a relatively short time scale, typically ranging from a few tens of nanoseconds to around 100 nanoseconds originated from light propagation in the TPC.
The shape of the $S1$ pulse is influenced by the PMT signal shaping, the decay time profile of singlet and triplet xenon dimers, and the propagation of photons within the TPC.
The ratio between the singlet and triplet dimer decays differs slightly between electronic recoils (ERs) and nuclear recoils (NRs). 
In ER events, where the incoming particle interacts with the xenon shell electrons, this ratio tends to be slightly higher compared to NR events~\cite{kubota1978evidence, hitachi1983effect}, where the interaction occurs with the xenon nucleus.
However, the difference in pulse shapes between ER and NR events is obscured as the consequence of the photon propagation effects, such photon Rayleigh scattering as well as the reflection on the liquid surface and TPC wall.

\begin{figure}[htp]
    \centering
    \includegraphics[width=0.8\textwidth]{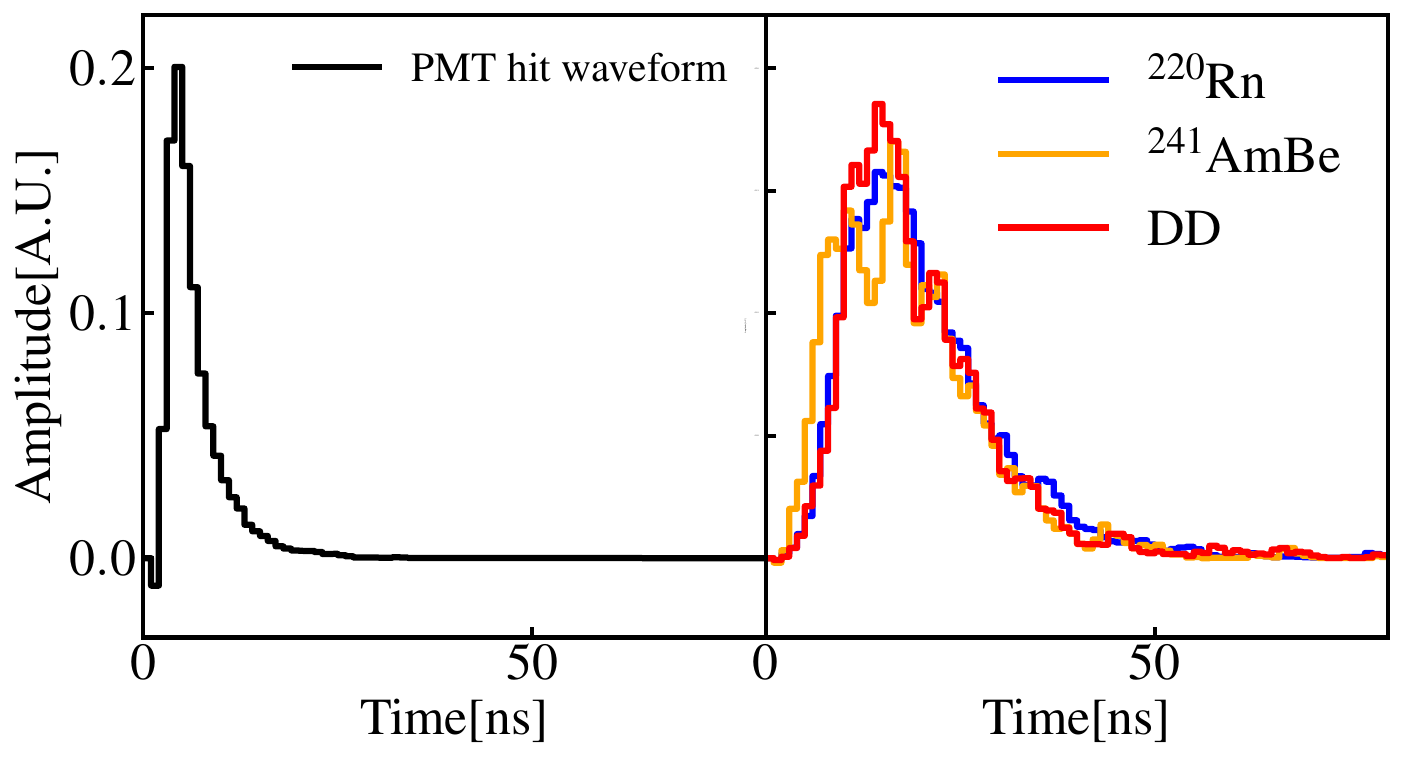}
    \caption{
    The single PE waveform and the time distributions of the $S1$ hits are shown in the left and right panels, respectively.
    The blue, orange, and red solid lines in the right panel represent the distributions from $^{220}$Rn, $^{241}$AmBe, and DD calibration data, respectively.
    }
    \label{fig:s1_time_profile}
\end{figure}

The simulation of the $S1$ waveform adopts a data-driven approach. 
In this approach, the $S1$ signals from $^{220}$Rn~\cite{ma2020internal} and neutron calibration data, including $^{241}$AmBe radioactive source and Deuteron-Deuteron (DD) neutron generator, are utilized to simulate the $S1$ waveforms corresponding to ERs and NRs, respectively.
$S1$s with the charge in  2 to 200 photoelectrons (PE) and 2 to 150\,PE, respectively, are selected as the ER and NR $S1$ pools.
Within these charge levels, the $S1$ waveforms consist of single-photon hits that are distributed among multiple PMTs and are distinguishable in time.
The time profile of a single-photon hit on a PMT is depicted in the left panel of Fig.~\ref{fig:s1_time_profile}. 
To create a simulated $S1$ waveform that shared the same event position as the pooled $S1$ (transverse distance $<$5\,cm and vertical distance $<$8\,cm), a fraction of the hit waveforms of the pooled $S1$ are randomly selected to form a simulated $S1$ with reduced charge.
This data-driven approach naturally takes into account the position-dependent characteristics of the $S1$ pulse time profile resulting from light propagation.
Right panel of Fig.~\ref{fig:s1_time_profile} displays the hit time distributions of $S1$ signals in the ER and NR calibration data from various $Z$ positions.
The difference in shape of waveforms arises from the difference of spatial position distributions in ER and NR calibration data.

\subsection{S2 pulse}

\begin{figure}[htp]
    \centering
    \includegraphics[width=0.8\textwidth]{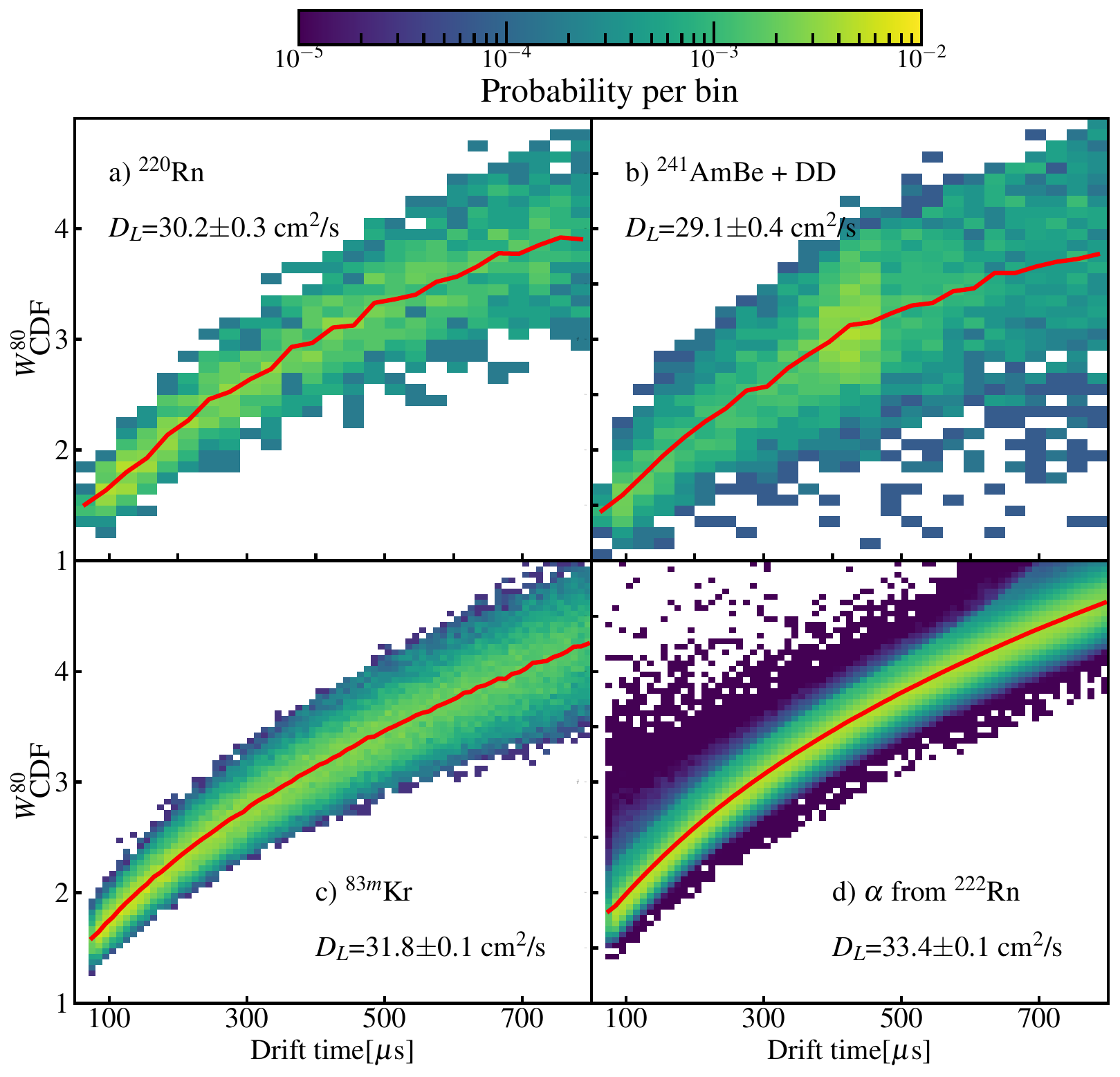}
    \caption{
    The normalized distributions of the $W^{80}_{\textrm{CDF}}$ over the drift time for four types of data: a) the $^{220}$Rn calibration, b) the neutron calibration using $^{241}$AmBe and DD, c) the $^{83m}$Kr calibration, and d) the $\alpha$ events originated from $^{222}$Rn impurities.
    The red solid lines represent the means of the distributions.
    The $D_L$ values obtained from a fit to the means using Eq.~\ref{eq:diffusion} are given in each panel.
    }
    \label{fig:alpha_kr_diffusion_coeff}
\end{figure}

\begin{figure}[htp]
    \centering
    \includegraphics[width=0.8\textwidth]{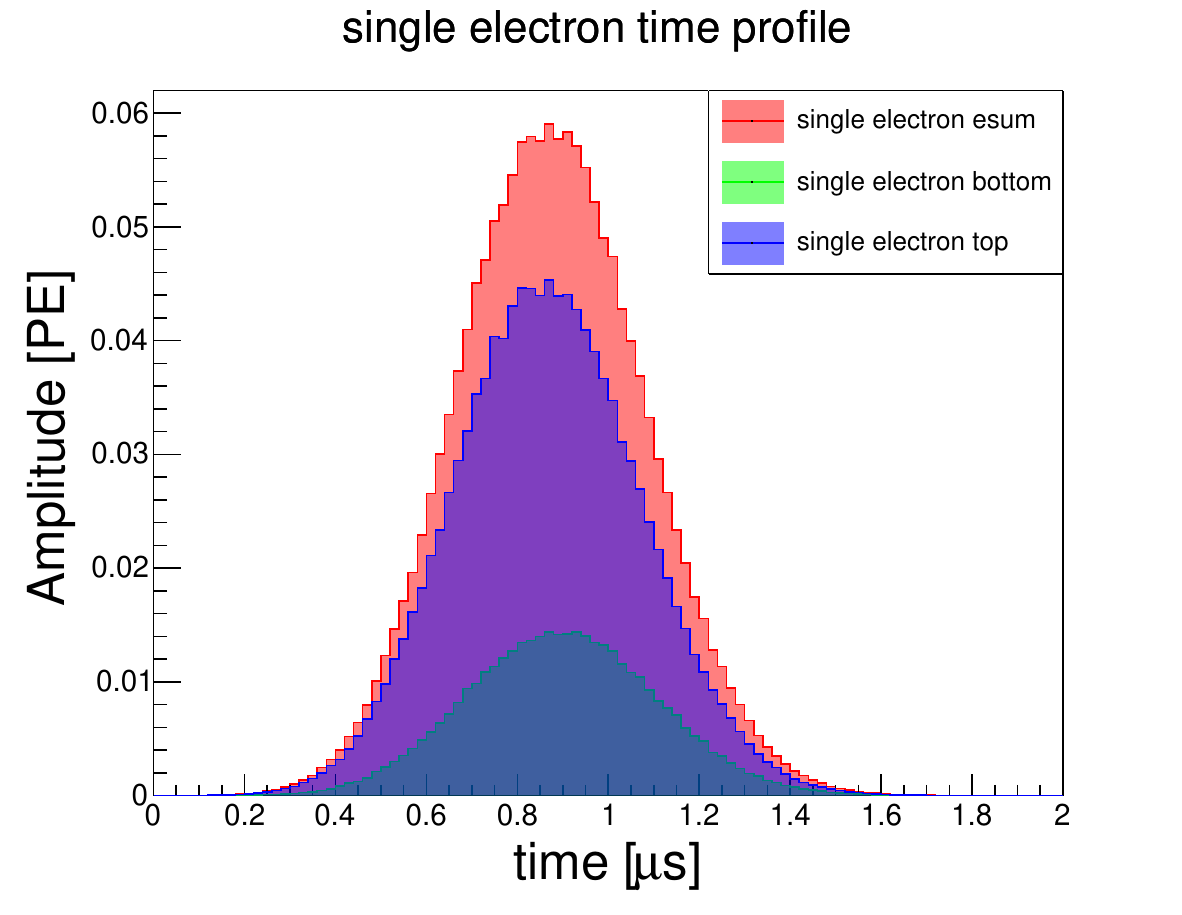}
    \caption{
    Average single electron waveforms.
    }
    \label{fig:se_time_profile}
\end{figure}

The $S2$ signal corresponds to the detection of proportional scintillation light emitted by the drifted electrons when they reach the gaseous xenon layer between the anode and gate electrodes in the TPC detector.
The shape of the $S2$ pulse is predominantly determined by two factors: the diffusion of electrons during their drift from the interaction vertex to the gaseous xenon layer, and the subsequent travel within the gaseous layer. 
The longitudinal diffusion coefficient in liquid xenon has been experimentally measured to be approximately 50\,cm$^2$/s at an electric field of 500\,V/cm~\cite{njoya2020measurements}.
To obtain accurate $\textit{in-situ}$ values for the longitudinal diffusion coefficient, measurements were performed using different calibration sources. 
Specifically, the 41 keV gamma line from the $^{83m}$Kr calibration data, and the $\alpha$ decay events from $^{222}$Rn were utilized. 
Fig.~\ref{fig:alpha_kr_diffusion_coeff} illustrates the 80\%-CDF $S2$ width (defined as the length of time window that covers central 80\% of the $S2$ charge) over drift time distributions for these events, and the mean 80\%-CDF $S2$ widths as a function of drift time were used to calculate the respective longitudinal diffusion coefficients.
The longitudinal diffusion coefficient $D_L$ can be obtained by fitting the mean 80\%-CDF $S2$ widths (denoted as $\langle W_{S2}^{80} \rangle$) as a function of drift time by the relation:
\begin{equation}
   \langle W_{S2}^{80} \rangle = f^{80} \sqrt{2 D_L T + \sigma^2_0},
   \label{eq:diffusion}
\end{equation}
where $T$ is the drift time and $\sigma_0$ represents the standard deviation of photon hit time in a single electron (SE) waveform.
$\sigma_0$ reflects the travel time of the electron in the gaseous xenon layer, and depends on the gas gap and the electric field strength in the gap.
The factor $f^{80}=2.56$ is the conversion factor from the standard deviation to 80\%-CDF width assuming the $S2$ pulse shape is Gaussian.  
The best-fit $D_L$ using low-energy data ($^{220}\textrm{Rn}$, $^{241}$AmBe, and DD) is systematically lower than the results obtained using high-energy data ($^{83m}$Kr and $\alpha$ from $^{222}$Rn).
This is due to the statistical bias of the $S2$ pulse width once the $S2$ charge is only caused by a few electrons.
The variations observed in the diffusion coefficients obtained from high-energy $\alpha$ also suggest the potential presence of systematic effects in high-energy samples, such as PMT saturation.
Consequently, the WS employs the diffusion coefficient from $^{83m}$Kr calibration data. 
This coefficient has a good agreement in the width versus drift time distributions of the low-energy data, as illustrated by later text. 
The time profile of electron travel within the gaseous xenon can be determined by analyzing the SE waveforms, as depicted in Fig.~\ref{fig:se_time_profile}. Both the electric field and the decay characteristics of xenon dimers in the gaseous phase influence the observed time profile.

The SE waveforms from the data are reassembled to generate simulated $S2$ signals in the WS.
In order to ensure compatibility of the ($X$, $Y$) positions between the simulated $S2$ signals and the data, SE waveforms with reconstructed positions $R_0$ within a certain range, specifically within 4\,cm from the simulated position, are selected for the assembly process.
This value is determined to efficiently match the pattern of the $S2$ signals (see Sec.~\ref{sec:compare_to_data} for more detail).
The selected SE pulses' arrival times are shuffled to ensure that the simulated SE arrival times adhere to the diffusion principle, wherein the standard deviation of the arrival time, denoted as $\sigma_T$, must satisfy $\sigma_T = \sqrt{2 D_L T}$.

\subsection{PMT after pulsing}
\begin{figure}[htp]
    \centering
    \includegraphics[width=0.8\textwidth]{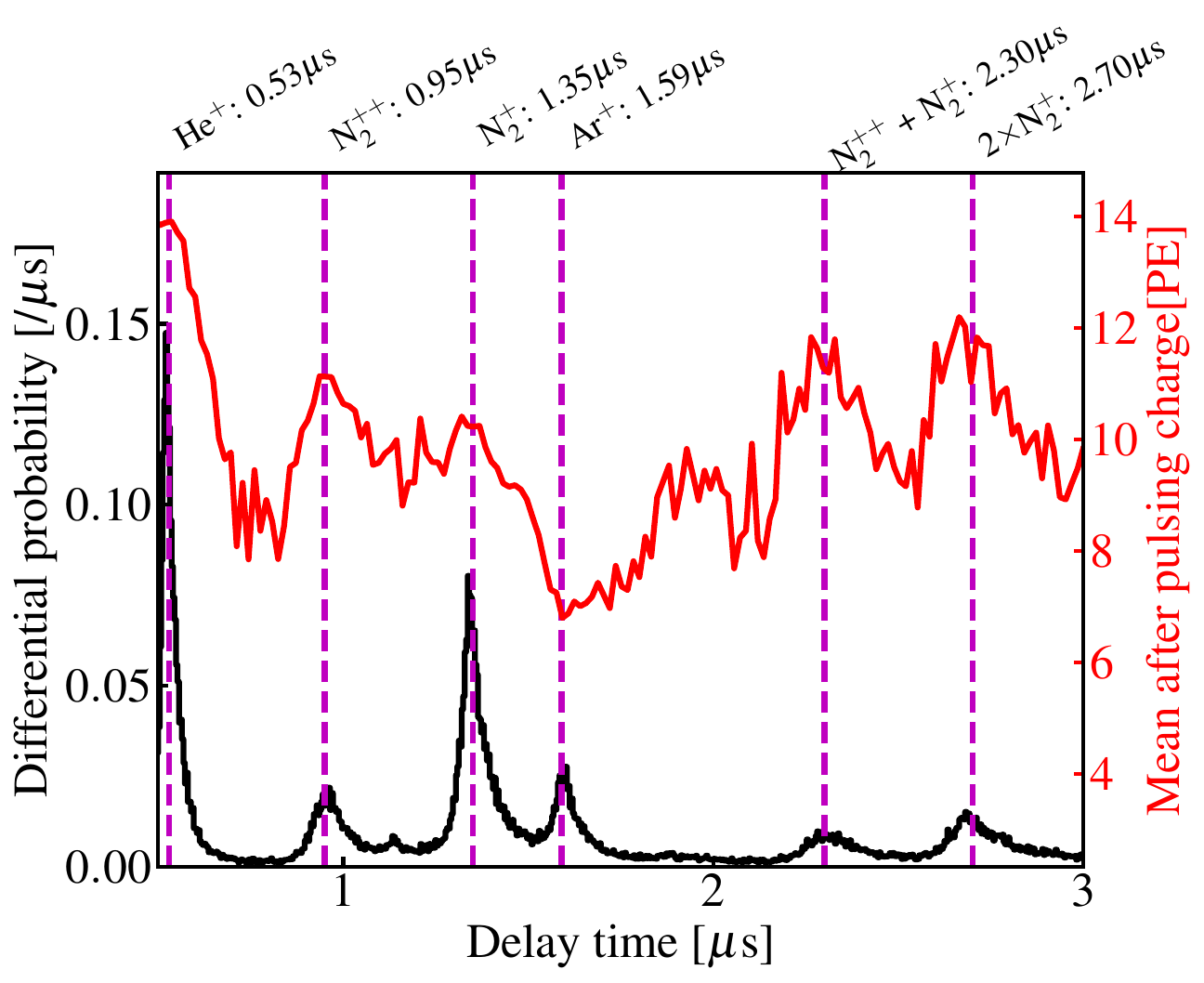}
    \caption{
    The differential probability of the PMT after pulsing as a function of delay time.
    The after pulsings caused by helium, nitrogen, and argon residual gases are most visible, and highlighted by the magenta dashed lines.
    There are also secondary after pulsing of nitrogen which is indicated by magenta dashed lines as well.
    The red solid line give the mean charge of after pulsing.
    }
    \label{fig:pmt_afterpulsing_profile}
\end{figure}

When a PMT detects light signals, residual impurities present within the PMT can become ionized due to the acceleration of photoelectrons (PEs)~\cite{li2016performance}.
These positively charged impurity ions then drift back towards the PMT's photocathode, causing the emission of additional PEs. 
As a result, small delayed signals may appear after the main $S1$ pulse.
The majority of impurities tend to be concentrated on the first dynode of the PMT, leading to characteristic and constant time delays for certain impurity ion in the PMT's response due to afterpulsing.
The magnitude of the time delay is primarily proportional to the square root of the atomic mass of the impurity.
Furthermore, the number of PEs that each impurity ion can generate is also dependent on the type of impurity.
In WS, the number of PMT afterpulsing hit and the charge of each hit are sampled according to the average differential probability and mean charge of PMT afterpulsing, respectively, as a function of the delay time, shown in Fig.~\ref{fig:pmt_afterpulsing_profile}.
The cumulative probability of afterpulsing occurrence in the 3-inch PMTs utilized in the PandaX-4T experiment is approximately 3\%.


\subsection{Photoionization and delayed electrons}

\begin{figure}[htp]
    \centering
    \includegraphics[width=0.8\textwidth]{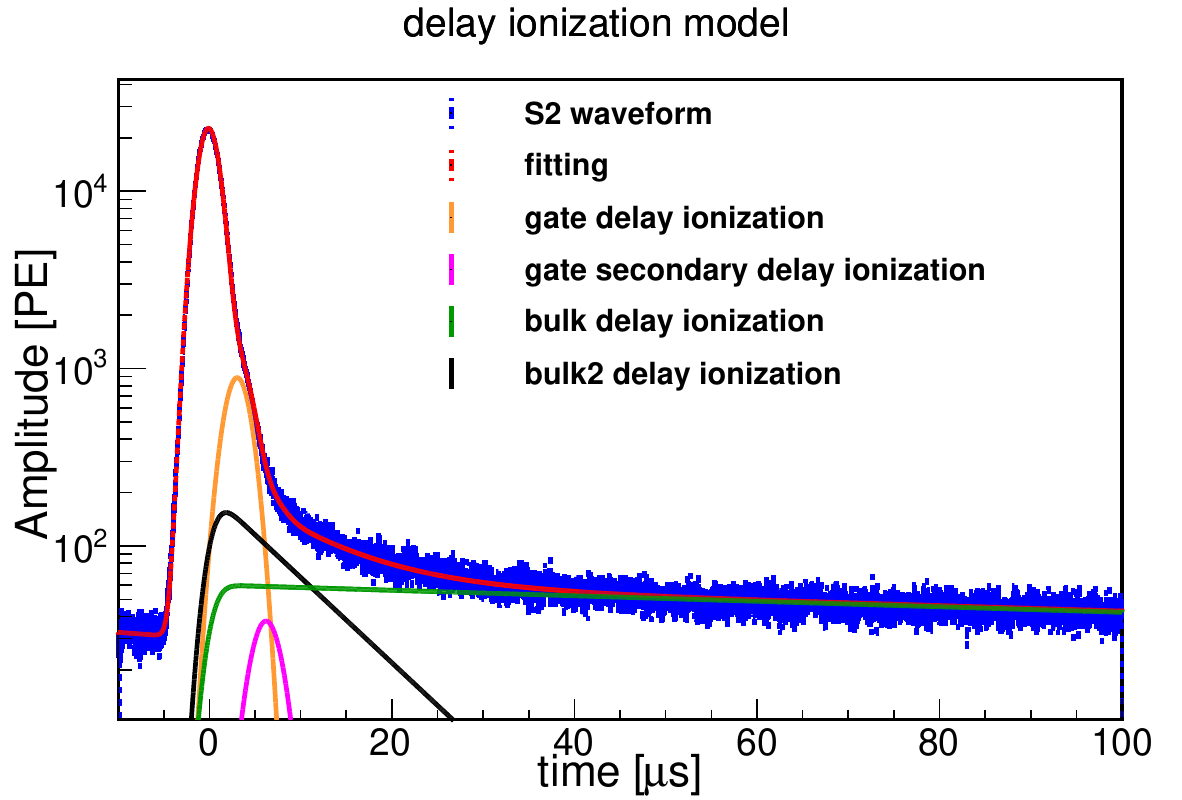}
    \caption{
    The average $S2$ waveform is displayed in blue solid line.
    The red solid line gives a fit to the average $S2$ waveform, which consists of the gate ionization (orange solid line) and the delayed electron contribution from the bulk of LXe (black and green solid lines).
    The gate photoionization is modeled as Gaussian, and the bulk photoionization is modeled as two exponentials.
    The $S2$ light of gate photoionization can also further generate photoionization (secondary gate photoionzation), which is included in the plot and shown as the magenta solid line.
    }
    \label{fig:s2_after_pulse}
\end{figure}

In several LXe-TPC detectors, it has been observed that additional $S2$ signals appear subsequent to a large $S2$ signal~\cite{aprile2014observation, aprile2022emission, akerib2020investigation}, commonly referred to as delayed $S2$ or delayed electrons. 
The appearance of these delayed signals is attributed to various factors. 
One significant factor is the photoionization of the electrode metal. 
The electrons resulting from this photoionization process exhibit distinct drift times, as depicted in Fig.~\ref{fig:s2_after_pulse}.
Delayed electrons can also arise from the photoionization of impurities uniformly distributed within the detector. 
It is also speculated that impurities present in the LXe may "capture" and subsequently release drifting electrons~\cite{akerib2020investigation}, leading to delayed $S2$ signals. 
Other processes, such as electron trapping at the liquid-gas interface~\cite{sorensen2018two} and spontaneous electron emission from the electrode~\cite{tomas2018study, bailey2016dark}, have also been proposed as potential contributors to the delayed $S2$ phenomenon.
In the PandaX-4T experiment, a data-driven approach is employed to model the probability of delayed electron generation. 
By stacking selected $S2$ waveforms with a fixed reference time (e.g., the start of the waveform), the resulting stacked waveform is analyzed to give the production probability, as shown in Fig.~\ref{fig:s2_after_pulse}. 
The correlation between the probability of the delayed electron production and the corresponding delay times is modeled empirically using a combination of two Gaussians and two exponential distributions.
The Gaussian components represent delayed electrons originating from the gate and cathode electrodes, respectively. 
The mean delay time for the gate delayed electrons, extracted from the fit to the stacked waveform, is 3.2\,$\mu$s. 
These values align with the expected behavior, assuming the liquid-gas interface is 5-8\,mm above the gate.
The delayed electrons from cathode are negligible compared to other components.
The two exponential distributions are found to adequately model delayed electrons resulting from impurity photoionization and electron delays caused by impurity or liquid surface trapping.
The probabilities of generating one delayed electron per detected $S2$ photon are determined as 0.24\% and 0.15\%, respectively, for the gate photoionization and other aforementioned effects.
In the WS, number of delayed electrons are sampled based on these probabilities and the delay time distributions shown in Fig.~\ref{fig:s2_after_pulse}.
The secondary photoionization caused by the primary photoionization is negligible and not implemented in WS.
Further adjustments to the photoionization probability are made to account for the presence of small $S2$ signals in the $S2$ waveform samples obtained from real data.

\subsection{Noise and dark counts}

\begin{figure}[htp]
    \centering
    \includegraphics[width=0.6\textwidth]{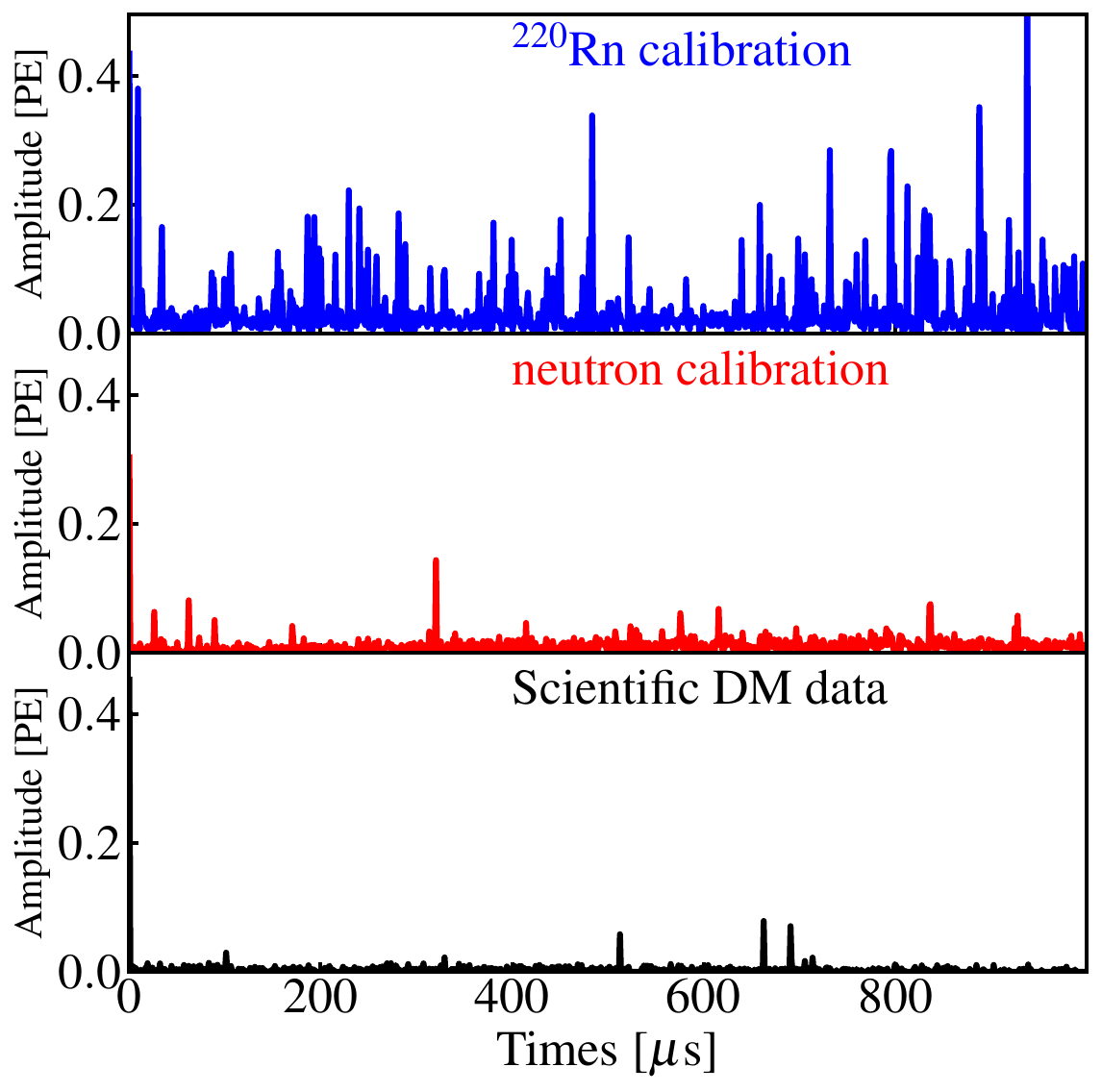}
    \caption{
    Sample noise waveforms from $^{220}$Rn calibration data (top), the neutron calibration data (middle), and DM data (bottom).
    }
    \label{fig:noise_wfs}
\end{figure}

Apart from the aforementioned effects, it is important to consider the presence of noise, including the spurious waveforms, and dark counts from PMTs, as they can potentially overshadow the small $S1$ signals and lead to incorrect pairing. 
To account for these sources of noise and dark counts, they are incorporated into the WS by stacking real segmented waveforms on top of the simulated waveform. 
These noise segments are randomly sampled from the 2-ms window preceding the identified $S1$s.
The level of noise and the rate of dark counts appear to have a correlation with the type of run being conducted.
Fig.~\ref{fig:noise_wfs} displays selected segmented waveforms from scientific runs for DM search, ER calibration, and NR calibration, respectively.


\section{Comparison to data}
\label{sec:compare_to_data}

To evaluate the performance of the WS, a comparative analysis is carried out by examining the distributions of key variables between the simulated samples and the experimental data. 
These key variables are derived by subjecting the simulated waveforms to the same data processing algorithm employed in the PandaX-4T experiment.
The variables selected for comparison can be classified into four distinct categories, with respect to $S1$ pulse shape, $S2$ pulse shape, pattern, and waveform ``dirtiness''. 
The description and exact definitions of these variables can be found in Table~\ref{tab:1}.

\begin{table}[htp]
    \centering
    \small
    \begin{tabular}{p{0.2\textwidth}|p{0.08\textwidth}|p{0.5\textwidth}|p{0.12\textwidth}}
    \hline\hline
    Category            &
    Symbol              & 
    Description         & 
    reference           \\
    \hline\hline
    \multirow{9}{*}{$S1$-related}       &
    $N^{\textrm{peak}}_{S1}$            &
    Number of peaks of the major $S1$.  &
    \multirow{9}{*}{Fig.~\ref{fig:s1_pulse_shape_comparison_er} and ~\ref{fig:s1_pulse_shape_comparison_nr}} \\
                                        &
    $Q_{S1}$                            &
    Total charge of the major $S1$.     &
                                        \\
                                        &
    $h_{S1}$                            &
    Pulse maximum height of the major $S1$. &
                                        \\
                                        &
    $W_{S1}$                            &
    The end time minus the start time of major $S1$ pulse.    &
                                        \\
                                        &
    $W_{S1}^{\textrm{FWHM}}$            &
    Half-height width of the major $S1$.    &
                                        \\
                                        &
    $W_{S1}^{\textrm{ten}}$                       &
    10\%-height width of the major $S1$.    &
                                        \\
                                        &
    $N^{\textrm{hit}}_{S1}$             &
    Number of hits for the major $S1$.  &
                                        \\
                                        &
    $N^{\textrm{cand}}_{S1}$            &
    Number of candidate $S1$s in the event.     &
                                        \\
                                        &
    $M^{\textrm{bot}}_{S1}$             &
    Charge on the most-fired bottom PMT for a major $S1$. &
                                        \\
    \hline
    \multirow{9}{*}{$S2$-related}       &
    $N^{\textrm{peak}}_{S2}$            &
    Number of peaks of the major $S2$.  &
    \multirow{9}{*}{Fig.~\ref{fig:s2_pulse_shape_comparison_er} and~\ref{fig:s2_pulse_shape_comparison_nr}} \\
                                        &
    $Q_{S2}$                            &
    Total charge of the major $S2$.     &
                                        \\
                                        &
    $W^{\textrm{FWHM}}_{S2}$            &
    Half-height width of the major $S2$.    &
                                        \\
                                        &
    $W^{\textrm{ten}}_{S2}$            &
    10\%-height width of the major $S2$.    &
                                        \\
                                        &
    $h_{S2}$                            &
    Pulse maximum height of the major $S2$. &
                                        \\
                                        &
    $R_{\textrm{pre}S2}$                            &
    Fraction of charge in pre-maximum-height window to total for the major $S2$. &
                                        \\
                                        &
    $N^{\textrm{hit}}_{S2} $            &
    Number of hits for the major $S2$.  &
                                        \\
                                        &
    $W^{80}_{S2}$                       &
    $S2$ width that contain 80\% charge. &
                                        \\
                                        &
    $\sigma^{\textrm{hit}}_{S2}$        &
    Standard deviation of the charges of hits among PMTs for the major $S2$. &
                                        \\
    \hline
    \multirow{6}{*}{Pattern-related}    &
    $A_{S1}$                            &
    Top-bottom asymmetry of the major $S1$. &
    \multirow{6}{*}{Fig.~\ref{fig:pattern_comparison_er} and ~\ref{fig:pattern_comparison_nr}} \\
                                        &
    $A_{S2}$                            &
    Top-bottom asymmetry of the major $S2$. &
                                        \\
                                        &
    $F_{S1}^{\textrm{maxq}}$                            &
    Fraction of charge on most-fired PMT over total charge of the major $S1$. &
                                        \\
                                        &
    $F_{S2}^{\textrm{maxq}}$                            &
    Fraction of charge on most-fired PMT over total charge of the major $S2$. &
                                        \\
                                        &
    $\sigma^{ch}_{S1}$                  &
    Standard deviation of the charges of fired PMTs for the major $S1$.                               &
                                        \\
                                        &
    $\sigma^{\textrm{CoG}}_{S2_b}$               &
    Root-mean-square of the hit PMT positions to the center-of-gravity reconstructed position for the major $S2$. &
                                        \\
    \hline 
    \multirow{4}{*}{Waveform ``dirtiness''}     &
    $\rho_{\textrm{preS1}}$               &
    Charge density before the major $S1$.   &
    \multirow{4}{*}{Fig.~\ref{fig:waveform_dirtiness_comparison_er} and ~\ref{fig:waveform_dirtiness_comparison_nr}} \\
                                        &
    $\rho_{S1-S2}$                      &
    Charge density inbetween the major $S1$ and $S2$. &
                                        \\
                                        &
    $\rho_{\textrm{postS2}} $                    &
    Charge density after the major $S2$. &
                                        \\
                                        &
    $F_{\textrm{S1-S2}}$                &
    Charge fraction of the major $S1$ plus $S2$ to the total charge in event waveform.                  &
                                        \\
    \hline\hline
    \end{tabular}
    \caption{
    \label{tab:1}
    List of key variables that are used in the comparison between data and WS.
    }
\end{table}

The comparison is conducted using both the $^{220}$Rn and neutron calibration data for ER and NR, respectively.
The data selections that are used for DM search~\cite{pandax4t_analysis_paper} are applied.
To mitigate the potential influence arising from correlations between the key variables and parameters such as $S1$ charge, $S2$ charge, and event position, it is necessary to ensure that the distributions of $S1$, $S2$, and position in both the WS and data samples are compatible.

\subsection{Comparison of $S1$ and $S2$ pulse shape related variables}

\begin{figure}[htp]
    \centering
    \includegraphics[width=0.8\textwidth]{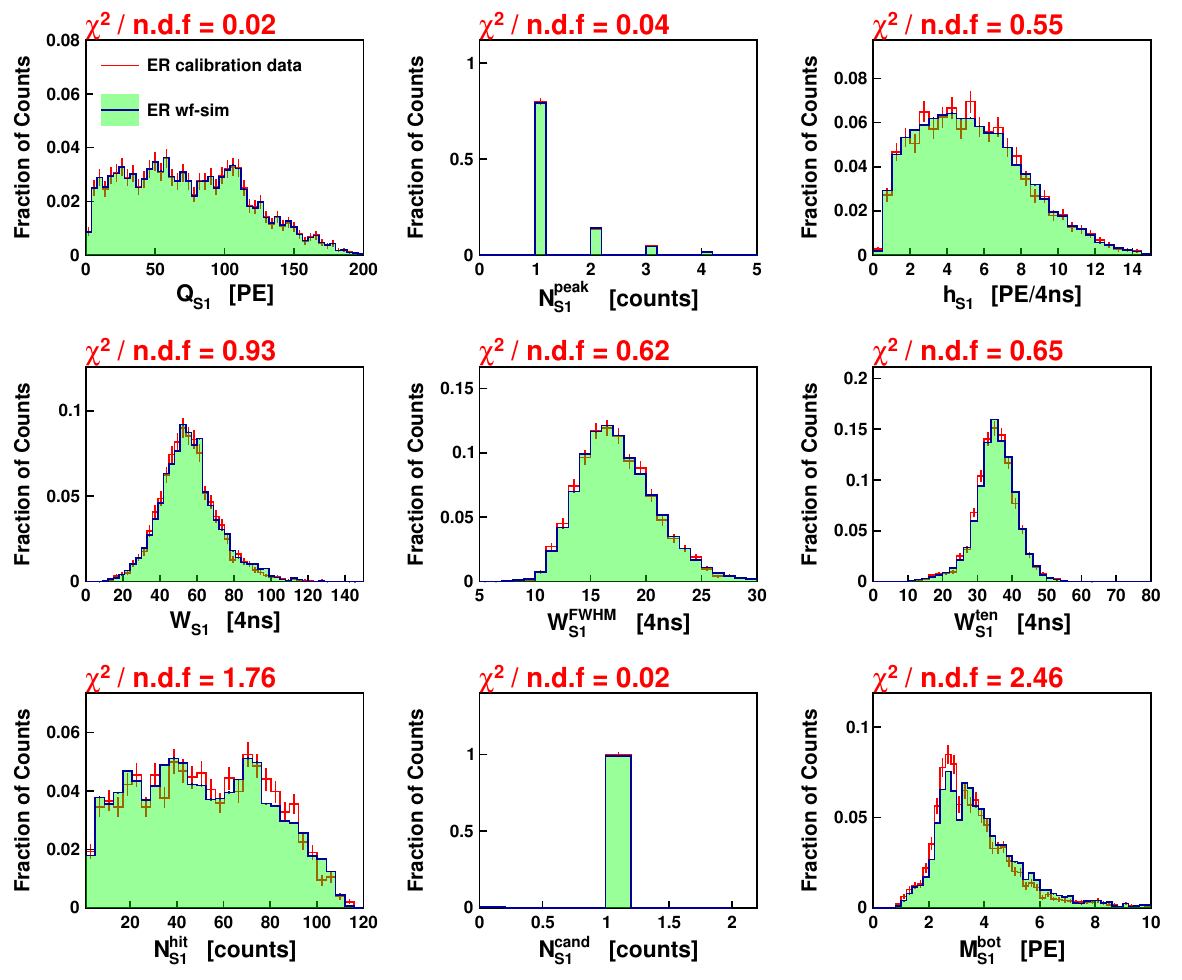}
    \caption{
    Comparisons of S1 key variables which are related to pulse shapes using ER calibration data.
    The red error bars represent the data distribution, and the shaded green histogram gives the distribution from the WS.
    The $\chi^2$ values divided by the degree of freedom are shown on top of each panel. 
    }
    \label{fig:s1_pulse_shape_comparison_er}
\end{figure}

\begin{figure}[htp]
    \centering
    \includegraphics[width=0.8\textwidth]{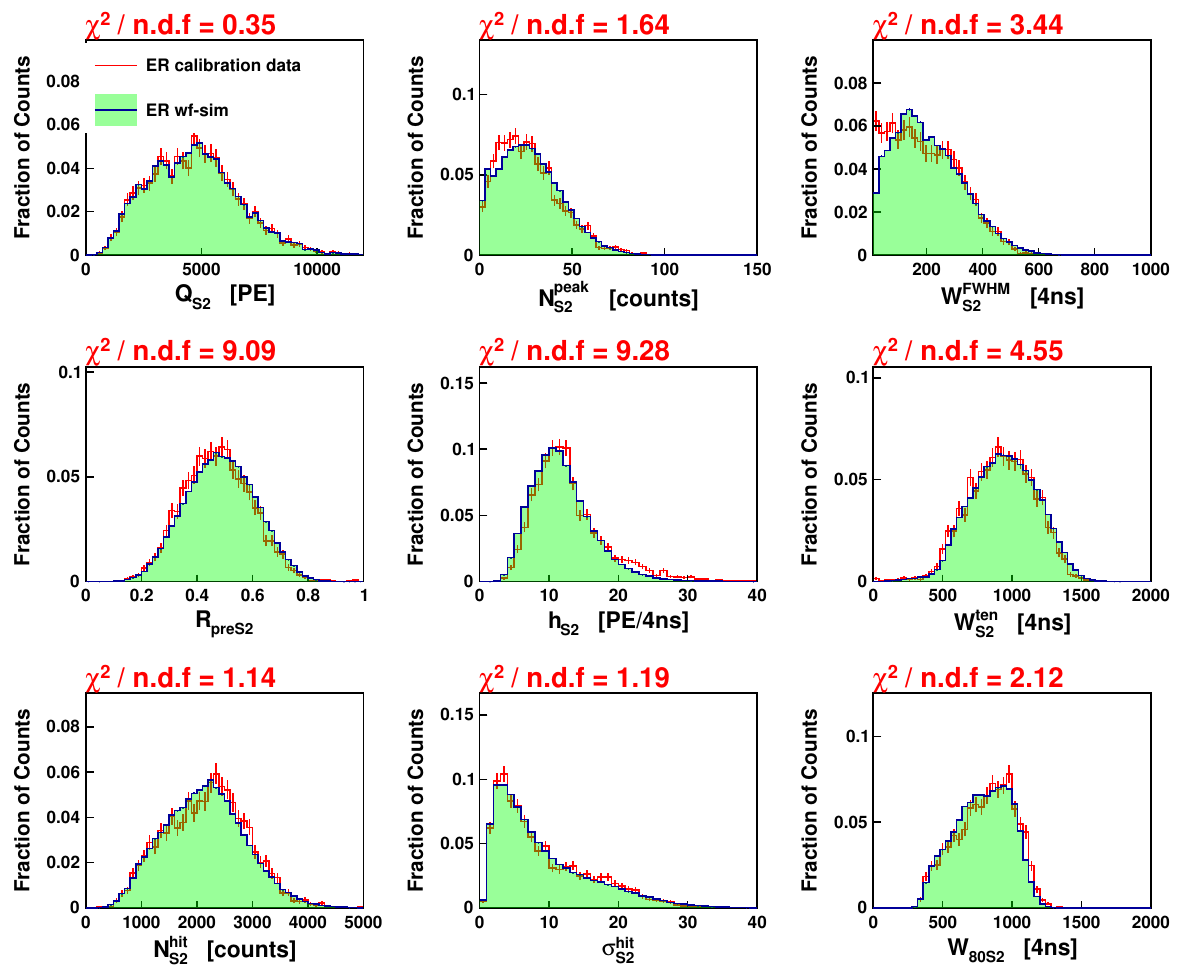}
    \caption{
    Comparisons of S2 key variables which are related to pulse shapes using ER calibration data.
    The red error bars represent the data distribution, and the shaded green histogram gives the distribution from the WS.
    The $\chi^2$ values divided by the degree of freedom are shown on top of each panel. 
    }
    \label{fig:s2_pulse_shape_comparison_er}
\end{figure}

\begin{figure}[htp]
    \centering
    \includegraphics[width=0.8\textwidth]{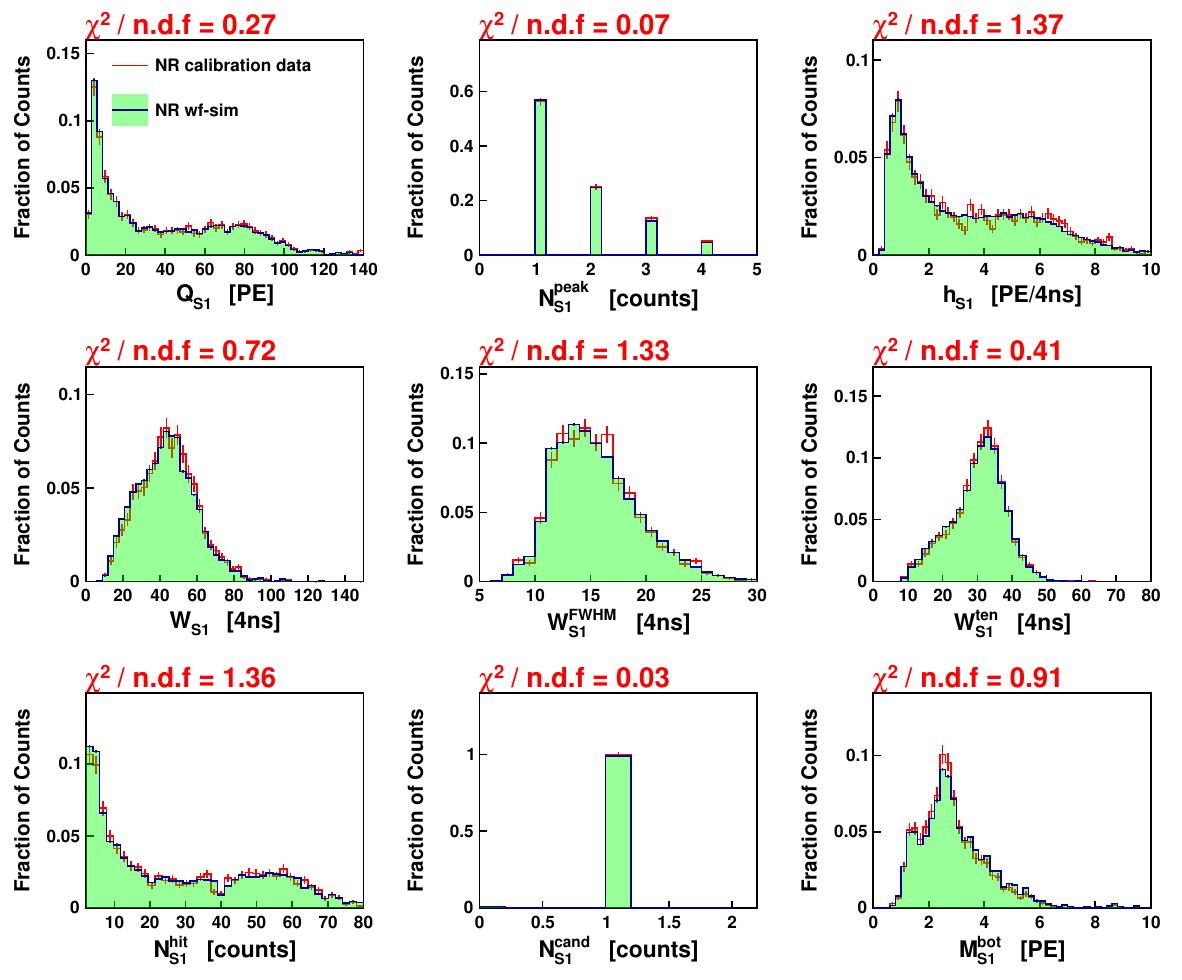}
    \caption{
    Comparisons of S1 key variables which are related to pulse shapes using NR calibration data.
    The red error bars represent the data distribution, and the shaded green histogram gives the distribution from the WS.
    The $\chi^2$ values divided by the degree of freedom are shown on top of each panel. 
    }
    \label{fig:s1_pulse_shape_comparison_nr}
\end{figure}

\begin{figure}[htp]
    \centering
    \includegraphics[width=0.8\textwidth]{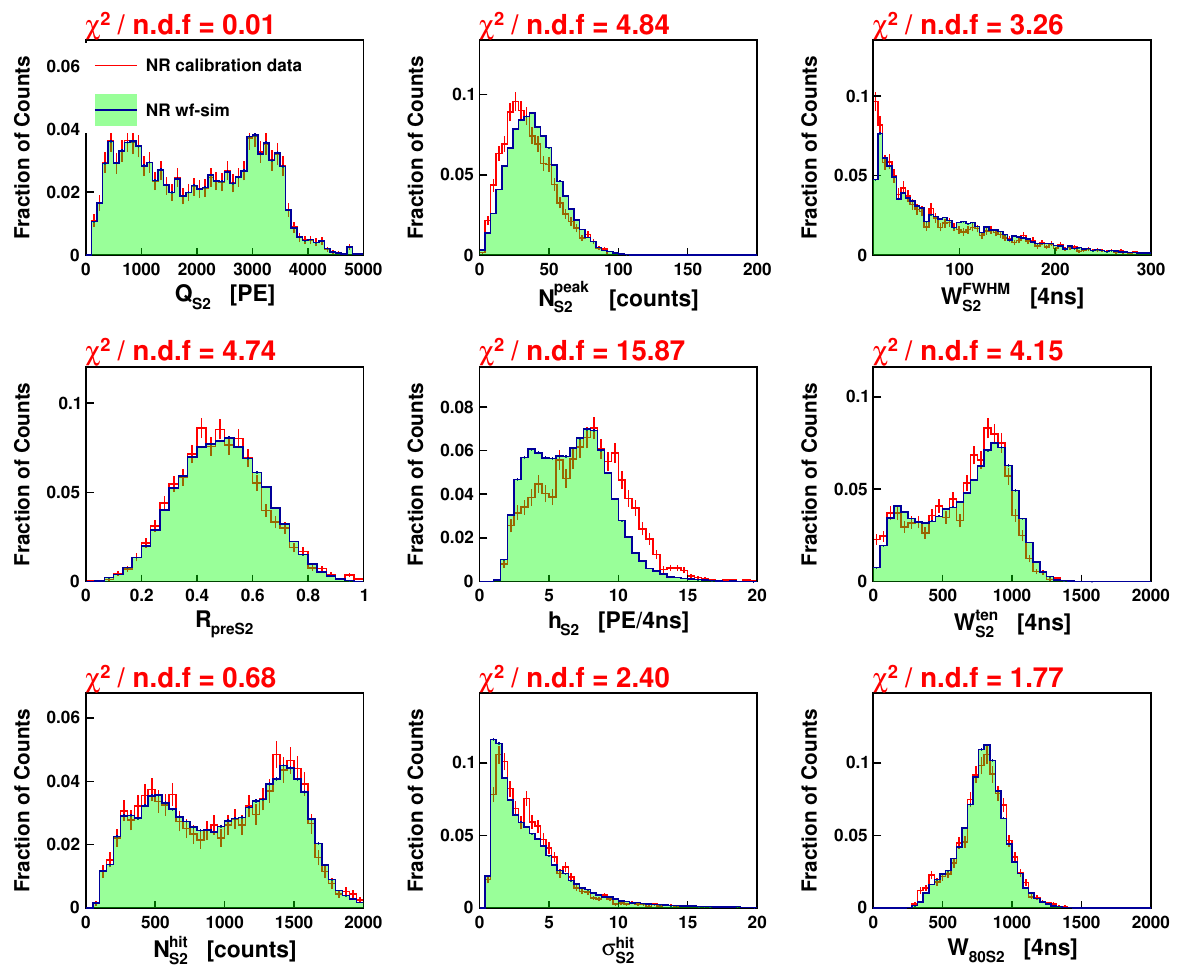}
    \caption{
    Comparisons of S2 key variables which are related to pulse shapes using NR calibration data.
    The red error bars represent the data distribution, and the shaded green histogram gives the distribution from the WS.
    The $\chi^2$ values divided by the degree of freedom are shown on top of each panel. 
    }
    \label{fig:s2_pulse_shape_comparison_nr}
\end{figure}

It is crucial to verify the fidelity of the WS by closely matching the pulse shapes of the simulated waveforms to those observed in real data. 
To validate this aspect, a comparison is conducted between the WS and data based on a set of selected variables, including pulse height, pulse width, number of hits, hit charge variations, and the like.
The comparisons are performed using ER and NR calibration data, respectively, and the chi-square values are provided in each plot as a measure of agreement.
Fig.~\ref{fig:s1_pulse_shape_comparison_er}, ~\ref{fig:s1_pulse_shape_comparison_nr},~\ref{fig:s2_pulse_shape_comparison_er}, and~\ref{fig:s2_pulse_shape_comparison_nr}  illustrate the comparisons for $S1$ and $S2$ pulses in both ER and NR calibration data. 
The majority of variables demonstrate good agreement between the WS and data, with only a few variables  exhibiting noticeable differences, including $h_{S2}$, $\sigma_{S2}^{\textrm{hit}}$, and $N_{S2}^{\textrm{peak}}$.
This is speculated to arise from the limited statistical significance of the calibration data comprising the $S1$ and SE pools in the WS. 
In addition, a relaxed selection criterion is employed for these $S1$ and SE events in order to augment the statistics, potentially introducing impurities into the collected data.

\subsection{Comparison of pattern related variables}
\begin{figure}[htp]
    \centering
    \includegraphics[width=0.8\textwidth]{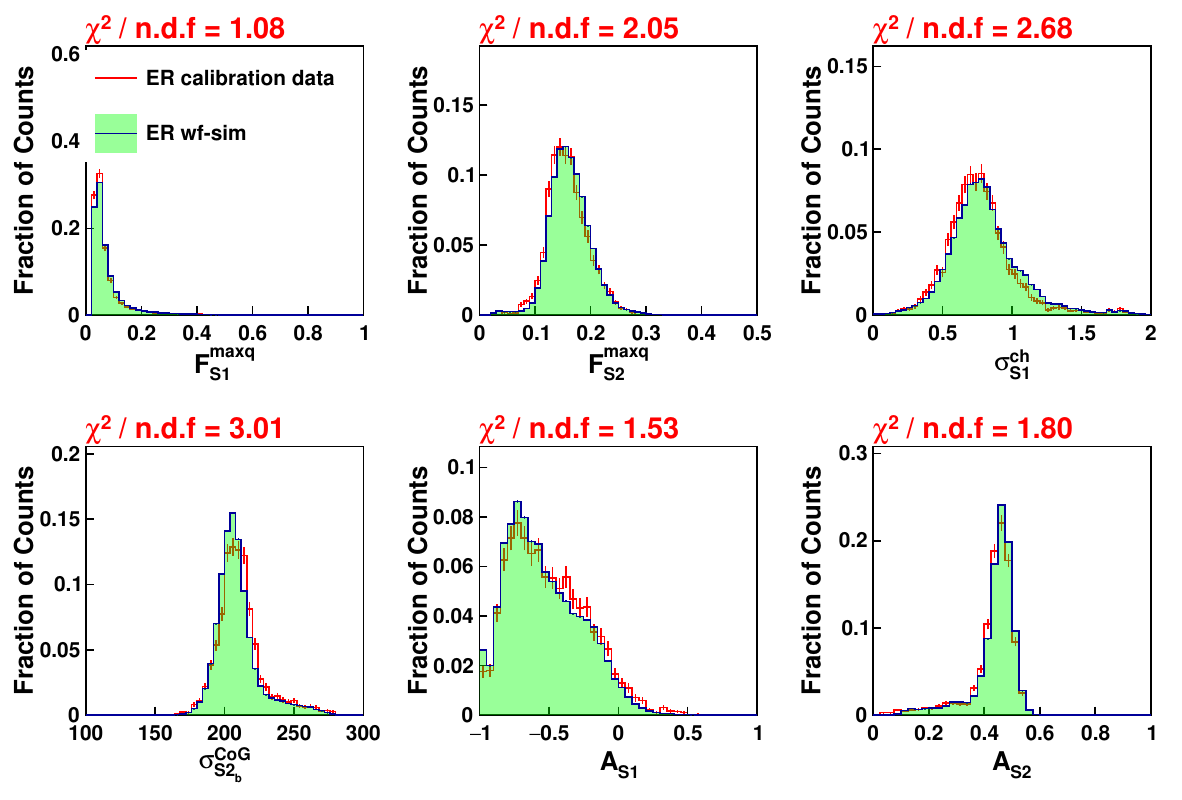}
    \caption{
    Comparisons of pattern-related key variables using ER calibration data.
    The red error bars represent the data distribution, and the shaded green histogram gives the distribution from the WS.
    The $\chi^2$ values divided by the degree of freedom are shown on top of each panel. 
    }
    \label{fig:pattern_comparison_er}
\end{figure}

\begin{figure}[htp]
    \centering
    \includegraphics[width=0.8\textwidth]{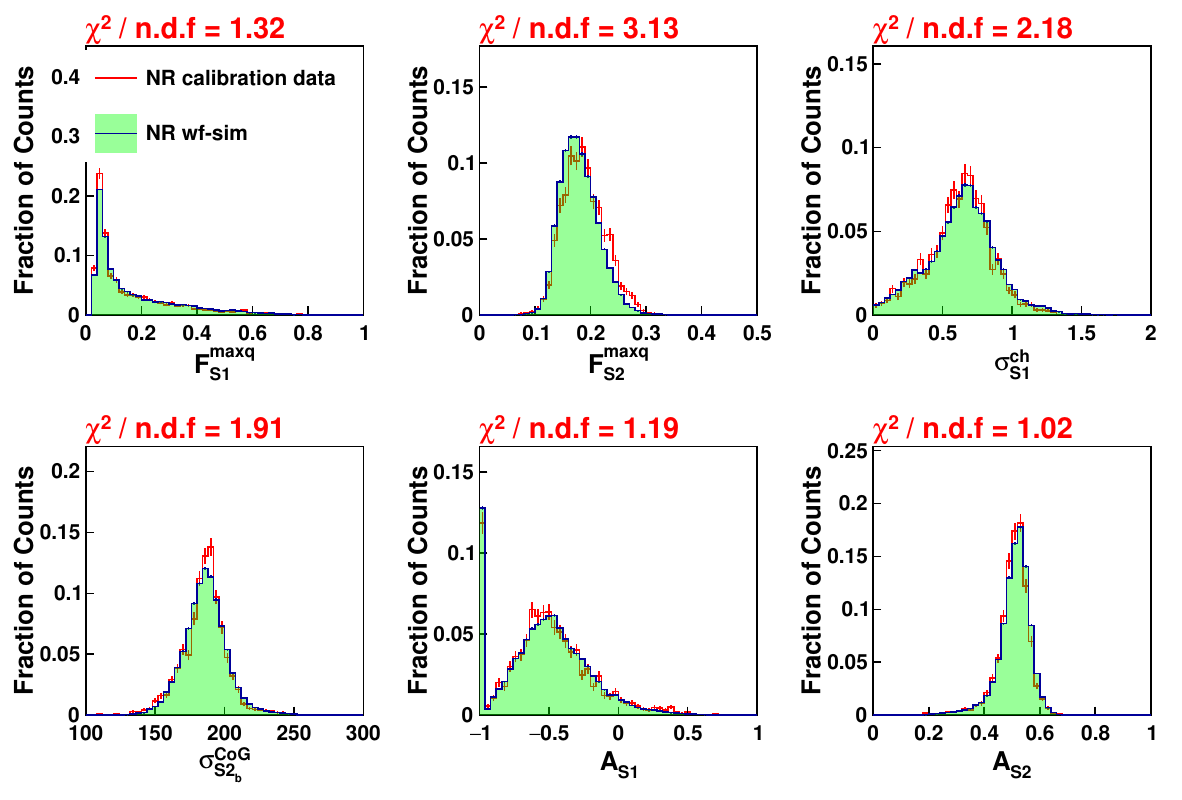}
    \caption{
    Comparisons of pattern-related key variables using NR calibration data.
    The red error bars represent the data distribution, and the shaded green histogram gives the distribution from the WS.
    The $\chi^2$ values divided by the degree of freedom are shown on top of each panel. 
    }
    \label{fig:pattern_comparison_nr}
\end{figure}

Not only the temporal features of the simulated pulses are of great importance, the pattern on PMTs plays also a crucial role.
Fig.~\ref{fig:pattern_comparison_er} and ~\ref{fig:pattern_comparison_nr} show the comparison of $S1$ and $S2$ pattern related variables between the WS and calibration data, including the top-bottom asymmetries for $S1$ and $S2$, the standard deviation of hit charges among fired PMTs, RMS of the hit PMT positions.
Most variables show good agreement with the exception for the fraction of $S2$ charge on the most-fired PMT over total $S2$ charge $F_{S2}^{\textrm{maxq}}$.
This is attributed to the inherent uncertainty in the reconstructed positions of the SEs within the simulation pool.
It is important to notice that the reconstructed positions of the SEs exhibit considerable statistical uncertainty due to the low $S2$ charge, particularly in proximity to the edge of the TPC.
Given that the actual positions of the SE blocks constituting the simulated $S2$ may not align with the expected position, it is comprehensible that discrepancies in $S2$ pattern-related variables between the data and WS arise.

\subsection{Comparison of waveform ``dirtiness'' related variables}

\begin{figure}[htp]
    \centering
    \includegraphics[width=0.8\textwidth]{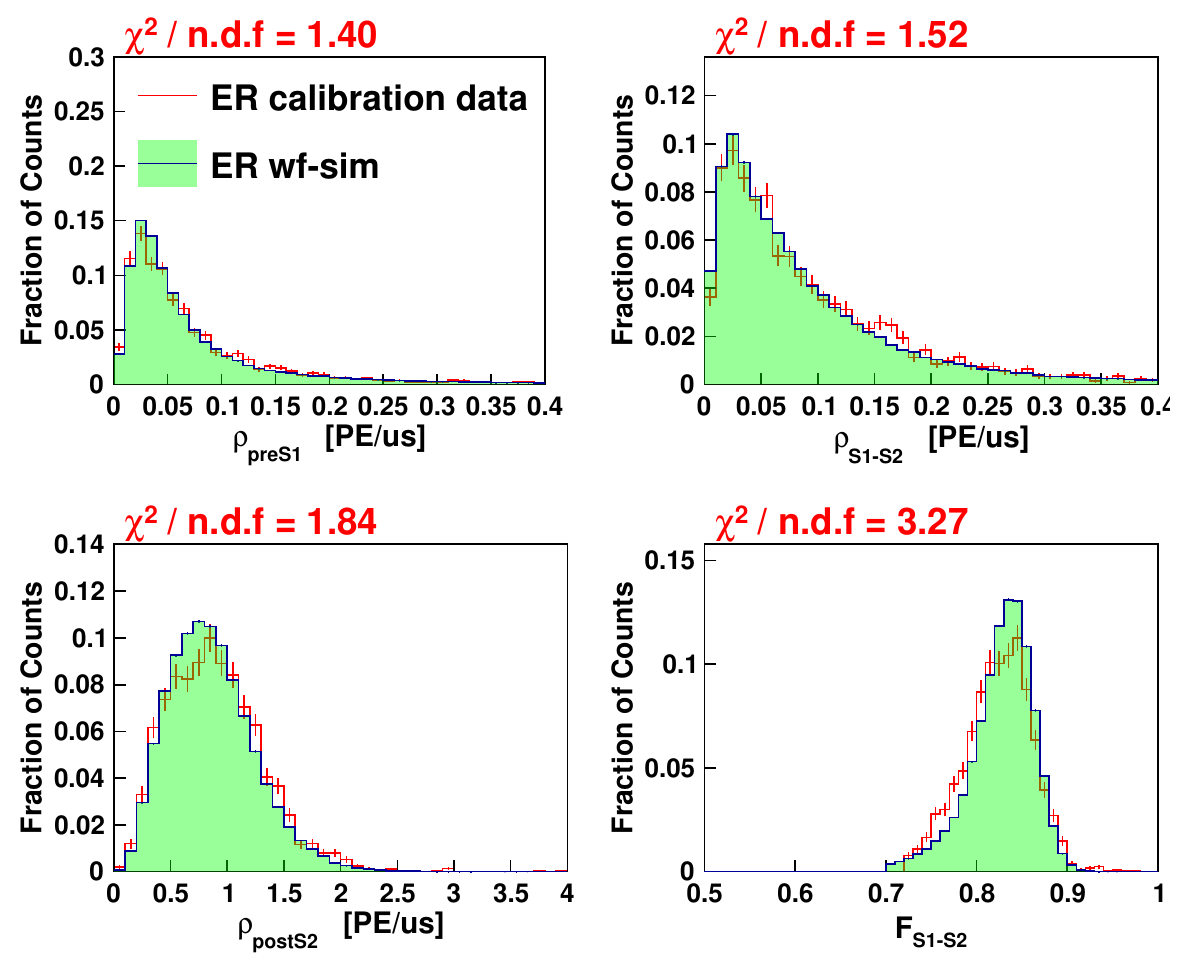}
    \caption{
    Comparisons of waveform ``dirtiness''-related key variables using ER calibration data.
    The red error bars represent the data distribution, and the shaded green histogram gives the distribution from the WS.
    The chi square values divided by the degree of freedom are shown on top of each panel. 
    }
    \label{fig:waveform_dirtiness_comparison_er}
\end{figure}

\begin{figure}[htp]
    \centering
    \includegraphics[width=0.8\textwidth]{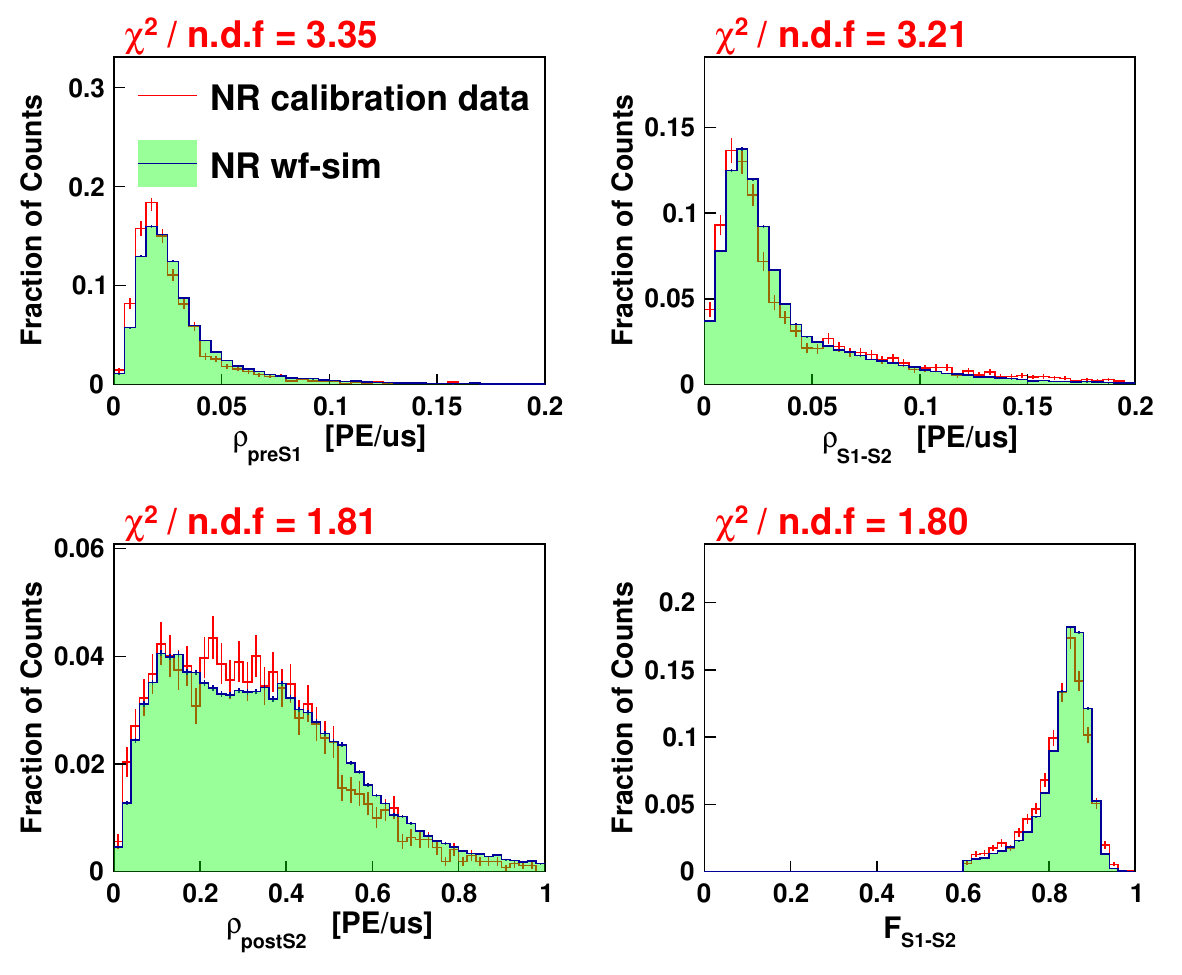}
    \caption{
    Comparisons of waveform ``dirtiness''-related key variables using NR calibration data.
    The red error bars represent the data distribution, and the shaded green histogram gives the distribution from the WS.
    The chi square values divided by the degree of freedom are shown on top of each panel. }
    \label{fig:waveform_dirtiness_comparison_nr}
\end{figure}

The presence of spurious noise in the waveforms can introduce incompatibility between the WS and the data. 
To evaluate this effect, direct comparisons are performed for several selected variables, as summarized in Table~\ref{tab:1} and depicted in Fig.~\ref{fig:waveform_dirtiness_comparison_er} and \ref{fig:waveform_dirtiness_comparison_nr}.
Initial results indicate slight deviations between the WS and the data, suggesting the need for adjustments in the production probability of the photoionization model within the WS. 
Subsequent modifications are made based on these findings, leading to a satisfactory improvement in the matching between the WS and the data.

\section{Summary and discussion}
\label{sec:summary_discussion}

We have presented a detailed overview of the waveform simulation framework developed for the PandaX-4T experiment. This WS framework has been extensively utilized to address various research objectives that require a sufficient statistical sample size, which may not be readily available in the experimental data.
The primary application of the WS is to generate synthetic samples for conducting studies related to reconstruction efficiency and bias, as referenced in~\cite{pandax4t_analysis_paper}. Additionally, the WS serves as a valuable resource for generating training samples for Boosted Decision Tree algorithms, as mentioned in~\cite{ma2023search}.
Through our analysis, we have demonstrated that the data-driven waveform simulation provides a comparable description of the experimental waveforms, particularly in terms of pulse shape, pattern, and the presence of spurious noise. This highlights the effectiveness and reliability of the WS in capturing important features of the observed data.
With more calibration data and understanding of the detector, we expect to further improve the WS, especially in terms of correlations between pulse shape and pattern, as well as its efficacy in high-energy regimes.
Additionally, we envision broadening the application of the WS to encompass the interpretation of a wider range of physical models.

\input{acknowledgement}


\input{main.bbl}

\end{document}

%% file: authorlist.tex

\author[16]{Jiafu Li}

\author[1]{Abdusalam Abdukerim}
\author[1]{Zihao Bo}
\author[1]{Wei Chen}
\author[1, 22]{Xun Chen}
\author[16]{Chen Cheng}
\author[17]{Zhaokan Cheng}
\author[14]{Xiangyi Cui}
\author[25]{Yingjie Fan}
\author[20]{Deqing Fang}
\author[20]{Changbo Fu}
\author[7]{Mengting Fu}
\author[4, 5, 6]{Lisheng Geng}
\author[1]{Karl Giboni}
\author[1]{Linhui Gu}
\author[8]{Xuyuan Guo}
\author[14]{Chencheng Han}
\author[1]{Ke Han}
\author[1]{Changda He}
\author[8]{Jinrong He}
\author[1]{Di Huang}
\author[21]{Yanlin Huang}
\author[1]{Junting Huang}
\author[1]{Zhou Huang}
\author[22]{Ruquan Hou}
\author[15]{Yu Hou}
\author[13]{Xiangdong Ji}
\author[15]{Yonglin Ju}
\author[1]{Chenxiang Li}
\author[8]{Mingchuan Li}
\author[8, 1]{Shuaijie Li}
\author[17]{Tao Li}
\author[3, 2]{Qing Lin\thanks{Corresponding author: qinglin@ustc.edu.cn}}
\author[1, 14, 22]{Jianglai Liu\thanks{Spokesperson: jianglia.liu@sjtu.edu.cn}}
\author[15]{Congcong Lu}
\author[11, 12]{Xiaoying Lu}
\author[7]{Lingyin Luo}
\author[3]{Yunyang Luo}
\author[1]{Wenbo Ma\thanks{Corresponding author: wenboma@sjtu.edu.cn}}
\author[20]{Yugang Ma}
\author[7]{Yajun Mao}
\author[1, 22]{Yue Meng}
\author[1]{Xuyang Ning}
\author[8]{Ningchun Qi}
\author[1]{Zhicheng Qian}
\author[11, 12]{Xiangxiang Ren}
\author[11, 12]{Nasir Shaheed}
\author[1]{Xiaofeng Shang}
\author[19]{Xiyuan Shao}
\author[4]{Guofang Shen}
\author[1]{Lin Si}
\author[8]{Wenliang Sun}
\author[13]{Andi Tan}
\author[1, 22]{Yi Tao}
\author[11, 12]{Anqing Wang}
\author[11, 12]{Meng Wang}
\author[20]{Qiuhong Wang}
\author[1, 23]{Shaobo Wang}
\author[7]{Siguang Wang}
\author[17, 16]{Wei Wang}
\author[15]{Xiuli Wang}
\author[1, 22, 14]{Zhou Wang}
\author[17]{Yuehuan Wei}
\author[16]{Mengmeng Wu}
\author[1]{Weihao Wu}
\author[1]{Jingkai Xia}
\author[13]{Mengjiao Xiao}
\author[16]{Xiang Xiao}
\author[14]{Pengwei Xie}
\author[1]{Binbin Yan}
\author[18]{Xiyu Yan}
\author[1]{Jijun Yang}
\author[1]{Yong Yang}
\author[1]{Yukun Yao}
\author[19]{Chunxu Yu}
\author[1]{Ying Yuan}
\author[20]{Zhe Yuan}
\author[1]{Xinning Zeng}
\author[13]{Dan Zhang}
\author[1]{Minzhen Zhang}
\author[8]{Peng Zhang}
\author[1]{Shibo Zhang}
\author[16]{Shu Zhang}
\author[1]{Tao Zhang}
\author[14]{Wei Zhang}
\author[11, 12]{Yang Zhang}
\author[11, 12]{Yingxin Zhang}
\author[14]{Yuanyuan Zhang}
\author[1]{Li Zhao}
\author[21]{Qibin Zheng}
\author[8]{Jifang Zhou}
\author[1, 22]{Ning Zhou}
\author[4]{Xiaopeng Zhou}
\author[8]{Yong Zhou}
\author[1]{Yubo Zhou}

\affil[1]{School of Physics and Astronomy, Shanghai Jiao Tong University, Key Laboratory for Particle Astrophysics and Cosmology (MoE), Shanghai Key Laboratory for Particle Physics and Cosmology, Shanghai 200240, China}
\affil[2]{State Key Laboratory of Particle Detection and Electronics, University of Science and Technology of China, Hefei 230026, China}
\affil[3]{Department of Modern Physics, University of Science and Technology of China, Hefei 230026, China}
\affil[4]{School of Physics, Beihang University, Beijing 102206, China}
\affil[5]{International Research Center for Nuclei and Particles in the Cosmos \& Beijing Key Laboratory of Advanced Nuclear Materials and Physics, Beihang University, Beijing 100191, China}
\affil[6]{School of Physics and Microelectronics, Zhengzhou University, Zhengzhou, Henan 450001, China}
\affil[7]{School of Physics, Peking University, Beijing 100871, China}
\affil[8]{Yalong River Hydropower Development Company, Ltd., 288 Shuanglin Road, Chengdu 610051, China}
\affil[9]{Shanghai Institute of Applied Physics, Chinese Academy of Sciences, 201800 Shanghai, China}
\affil[10]{Center for High Energy Physics, Peking University, Beijing 100871, Chin}
\affil[11]{Research Center for Particle Science and Technology, Institute of Frontier and Interdisciplinary Science, Shandong University, Qingdao 266237, Shandong, China}
\affil[12]{Key Laboratory of Particle Physics and Particle Irradiation of Ministry of Education, Shandong University, Qingdao 266237, Shandong, China}
\affil[13]{Department of Physics, University of Maryland, College Park, Maryland 20742, USA}
\affil[14]{Tsung-Dao Lee Institute, Shanghai Jiao Tong University, Shanghai, 200240, China}
\affil[15]{School of Mechanical Engineering, Shanghai Jiao Tong University, Shanghai 200240, China}
\affil[16]{School of Physics, Sun Yat-Sen University, Guangzhou 510275, China}
\affil[17]{Sino-French Institute of Nuclear Engineering and Technology, Sun Yat-Sen University, Zhuhai, 519082, China}
\affil[18]{School of Physics and Astronomy, Sun Yat-Sen University, Zhuhai, 519082, China}
\affil[19]{School of Physics, Nankai University, Tianjin 300071, China}
\affil[20]{Key Laboratory of Nuclear Physics and Ion-beam Application (MOE), Institute of Modern Physics, Fudan University, Shanghai 200433, China}
\affil[21]{School of Medical Instrument and Food Engineering, University of Shanghai for Science and Technology, Shanghai 200093, China}
\affil[22]{Shanghai Jiao Tong University Sichuan Research Institute, Chengdu 610213, China}
\affil[23]{SJTU Paris Elite Institute of Technology, Shanghai Jiao Tong University, Shanghai, 200240, China}
\affil[24]{School of Physics and Technology, Nanjing Normal University, Nanjing 210023, China}
\affil[25]{Department of Physics,Yantai University, Yantai 264005, China}


%% file: acknowledgement.tex
\section*{Acknowledgements}

This project is supported in part by grants from National Natural Science
Foundation of China (Nos. 12090060, 12090061, 12005131, 11905128, 11925502, 12222505, 11835005), and by the Office of Science and
Technology, Shanghai Municipal Government (grant No. 22JC1410100). We thank supports from Double First Class Plan of
the Shanghai Jiao Tong University. 
We also thank the sponsorship from 
the Hongwen Foundation in Hong Kong, Tencent
Foundation in China, and Yangyang Development Fund. Finally, we thank the CJPL administration and
the Yalong River Hydropower Development Company Ltd. for
indispensable logistical support and other help.